\documentclass[12pt,preprint]{aastex}
\usepackage{natbib}
\begin{document}

\title{Tadpole Galaxies in the Hubble Ultra Deep Field}

\author{Bruce G. Elmegreen}
\affil{IBM Research Division, T.J. Watson Research Center, 1101
Kitchawan Road, Yorktown Heights, NY 10598} \email{bge@watson.ibm.com}

\author{Debra Meloy Elmegreen}
\affil{Vassar College, Dept. of Physics \& Astronomy, Box 745,
Poughkeepsie, NY 12604} \email{elmegreen@vassar.edu}

\begin{abstract}
Tadpole galaxies have a head-tail shape with a large clump of star
formation at the head and a diffuse tail or streak of stars off to one
side. We measured the head and tail masses, ages, surface brightnesses,
and sizes for 66 tadpoles in the Hubble Ultra Deep Field (UDF), and we
looked at the distribution of neighbor densities and tadpole
orientations with respect to neighbors.  The heads have masses of
$10^7-10^8\;M_\odot$ and photometric ages of $\sim0.1$ Gyr for
$z\sim2$. The tails have slightly larger masses than the heads, and
comparable or slightly older ages.  The most obvious interpretation of
tadpoles as young merger remnants is difficult to verify. They have no
enhanced proximity to other resolved galaxies as a class, and the
heads, typically $<0.2$ kpc in diameter, usually have no obvious
double-core structure. Another possibility is ram pressure interaction
between a gas-rich galaxy and a diffuse cosmological flow. Ram pressure
can trigger star formation on one side of a galaxy disk, giving the
tadpole shape when viewed edge-on. Ram pressure can also strip away gas
from a galaxy and put it into a tail, which then forms new stars and
gravitationally drags along old stars with it.  Such an effect might
have been observed already in the Virgo cluster. Another possibility is
that tadpoles are edge-on disks with large, off-center clumps.
Analogous lop-sided star formation in UDF clump clusters are shown.
\end{abstract}

\section{Introduction}

Elongated galaxies with bright clumps at one end are visible in deep
field images taken with HST or from the ground.  \cite{vdb96} called
them ``tadpole'' galaxies. \cite{ab96} found that 13.4\%$\pm$1.6\% of
Hubble medium-deep-field galaxies showed tidal-like distortions or
tails, compared to 8.7\%$\pm$1.0\% locally. As part of a general survey
of young galaxy morphology, we compiled a catalog \citep{e05} of 97
tadpole galaxies in the ACS images of the Hubble Ultra Deep Field
\citep[UDF;][]{beck06}. We also determined photometric redshifts for
these and other types larger than 10 pixels in diameter \citep{e07}. In
the redshift range from 1 to 4, tadpoles represented 10\% of the total
count of such galaxies, comparable to the fraction of chain galaxies
(13.4\%) and double galaxies (16.7\%), smaller than the clump-cluster
fraction (33.4\%) and spiral galaxy fraction (21.4\%), and larger than
the elliptical galaxy fraction (4.6\%). The spirals and ellipticals
could be partially obscured by bandshifting, considering that these
more conventional types tend to be intrinsically redder than the clumpy
types.

\cite{rh05} studied a galaxy somewhat like a tadpole, UDF 5225, which
they considered to have a core and a plume component, like a tadpole
head and tail. We classified UDF 5225 as a chain galaxy, because the
``plume'' has three clumps in it. Rhoads et al. measured the spectrum
throughout and found continuum and L$\alpha$ emission at redshift
$z=5.4$.  They concluded that the galaxy was a merger with star
formation in the core and in knots along a tidal tail.

\cite{straughn06} and \cite{win06} identified 165 tadpole galaxies in
the UDF using an automated search algorithm. They showed that the
photometric redshift distribution for tadpoles was about the same as
that of field galaxies, with tadpoles representing 6\% of the total.
Following simulation results by \cite{spring05a,spring05b}, who found a
tadpole phase 0.7-1.5 Gyr after a merger and the epoch of peak black
hole accretion $\sim1$ Gyr after that, Windhorst et al. compared the
tadpole population to UDF galaxies with some AGN-type variability. They
saw no overlap between the two populations, and suggested that this was
consistent with the Springel et al. prediction about the relative
timing of different phases. Straughn et al. also noted that every
galaxy would have to undergo 10 to 30 such mergers since redshift
$\sim7$ to account for the tadpole fraction, given the relatively short
time spent in this phase. They suggested that since tadpole shapes
require a certain type of merger, the actual number of strong
interactions or mergers per galaxy would have been larger.

\cite{rawat07} studied 39 Luminous Compact Galaxies that included 3
tadpoles with stellar $\log M/M_\odot=10.23$, 9.71, and 10.67. They
noted that the tadpole heads were bluer than the tails, and suggested
that this whole class of luminous galaxies is evolving toward
intermediate mass spirals, with major mergers playing an important role
in the observed morphology.  Mergers were inferred primarily from the
presence of multiple cores with similar luminosities in the rest-frame
B-band, or the presence of double red nuclei. One of their tadpoles has
a double-core head.

\cite{dem06a} show a tadpole galaxy in their Figure 16 and note that
many compact galaxies in their UV-selected sample have tadpole
morphologies. \cite{dem06b} studied 268 UV-bright sources in parallel
WFPC2 fields of the Great Observatories Origins Deep Survey
\citep[GOODS;][]{gia04} and visually classifying 93 with starburst
SEDs. They found that 36\% of the starbursts are tadpoles and 50\% of
the starbursts have another galaxy within a $5^{\prime\prime}$ area
(which corresponds to a 20 kpc radius at their average $z=1.5$). The
half-light radii of the tadpoles was $1.6\pm0.4$ kpc. Several are shown
in their Figures 11, and 12.

Here we use the ACS and NICMOS UDF fields to measure the head and tail
masses, ages, surface densities, and sizes for 66 tadpoles in our
morphology catalog. We find that tadpole heads are usually single at
the resolution of the ACS. They look like clumps in clump-cluster
galaxies, i.e., they are low-mass and young compared to whole L$^*$
galaxies.  Note that tadpoles in our survey are defined only by the
presence of a bright clump of light at one end and a smooth, somewhat
linear, tail extending away from the clump; they are not defined to
have single-component heads. We discuss examples of tadpole heads with
substructure in Section \ref{comp}; these could be mergers or they
could be complex structures from star formation in the heads.

We also look for an excess of neighbors around tadpoles. We compare the
projected densities of galaxies around them with the projected
densities of galaxies around random field galaxies in the UDF. No
difference is evident, nor is there a consistent orientation of
tadpoles relative to near-neighbor directions. A few tadpoles have
double heads (examples are shown), but generally there is little
off-axis structure, duplicity, or other evidence for mergers or
interactions. This lack of evidence is consistent with the recent
suggestion that galaxy mergers are less important for building galaxies
than formerly thought. Major mergers thicken stellar disks in
unacceptable ways \citep{be09,bem09}, and they are not essential for
building galaxies in the presence of cold flows
\citep{ocv08,dekel09,ager09,ker09,broo09}.

Other possible origins for tadpole structure are gas stripping or
star-formation triggering by ram pressure. The tail could be
star-forming gas that was stripped from the head and accompanied by
entrained older stars, or the head could be the leading surface of a
low-mass disk that was compressed into star formation. Ram-pressure
stripping and triggering are expected during galaxy motions through
dense cosmological gas flows in the neighborhood.

A related possibility is suggested by normal velocity gradients across
the tails of several tadpoles observed by \cite{forster09}. These
gradients suggest that the tadpoles are rotating disks like some
clump-clusters, but with a single large clump instead of a half-dozen
clumps. We suggest that the single clump could be the brightest part of
a ring or a bright clump near the edge of a stellar disk, and we show
examples of such galaxies.

We conclude with the suggestion that tadpoles could be a composite
population with several physical origins. Further studies of gas and
molecular abundances, internal dynamics, and galactic environment will
be necessary to understand the tadpole shape more fully.

In what follows, Section 1 discusses the data, Section 2 contains the
method of analysis for mass and age, Section 3.1 has the results for
mass, age, and surface density, 3.2 looks at companions, 3.3 considers
internal dynamics, and 3.4 has two examples of clustered tadpoles
suggestive of environmental effects, as well as examples of what may be
face-on tadpoles in the lop-sided disk interpretation, A discussion of
possible models for tadpoles is in Section 4, and the conclusions are
in Section 5.

\section{Data}

The tadpole galaxies catalogued in \cite{e05} were studied for the
present paper. There were 97 tadpoles in that catalog, all larger than
10 pixels in diameter. Here we excluded from our photometric study the
cases that looked obviously like mergers, were contaminated by other
galaxies, or seemed to have spiral structure, which puts them in a
different morphology category.  Obvious mergers are those with
disconnected parts, multiple bright regions with clear separations,
multiple tails and other classical tidal features, and off-axis
emission. Objects with such features are considered to be mergers in
the classical sense. These features make them look different from the
simplest tadpoles, which have single bright heads and single tails, all
on approximately the same axis. We discuss in section \ref{comp}
tadpoles with structured heads that might be late-stage mergers,
although even these have single tails and little off-axis structure. We
also discuss in section \ref{comp} tadpoles with slightly curved tails.
These could be interacting tadpole galaxies with unknown companions
that accelerate the main tadpoles in the off-axis direction. They also
do not look like conventional mergers, so they are included in our main
study here. We discuss them specifically later as special cases in
order to consider the possibility that some tadpoles might be
interacting or mergers even though they do not look like it in the
conventional sense.

We used photometric redshifts from \cite{rafel09}, which includes
ground-based UV measurements, and excluded tadpoles that were not in
the Rafelski et al. compilation. Other tadpoles were excluded for
faintness, particularly in B-band for high-$z$ dropouts. For the
remaining sample of 66 tadpoles, the AB magnitudes and number of pixels
were determined for the heads and for the prominent regions of the
tails using the four ACS passbands, $B_{435}$, $V_{606}$, $i_{775}$,
and $z_{850}$. Boxes around the objects were defined using the IRAF
task $\it imstat$. Typically the object boundaries were at a level
about 10$\sigma$ above the noise. Sky was not subtracted because it is
essentially zero.

NICMOS observations of the UDF \citep{thomp05} also included tadpole
galaxies. For 37 cases with Rafelski et al. redshifts, we measured the
J and H-band magnitudes of the heads, along with the $B_{435}$,
$V_{606}$, $i_{775}$, and $z_{850}$ ACS magnitudes of the same heads on
images convolved to the same resolution as NICMOS.

Zeropoint conversions for each filter were taken from the online
handbooks. Magnitude measurement errors are estimated to be about 0.1
mag. Boxes rather than circles were used to define magnitudes because
the regions were sometimes elongated.  The same box position was used
for each passband. Head and tail colors typically varied by less than
0.05 mag for different box placements or sizes.

\begin{figure}\epsscale{1.}
\plotone{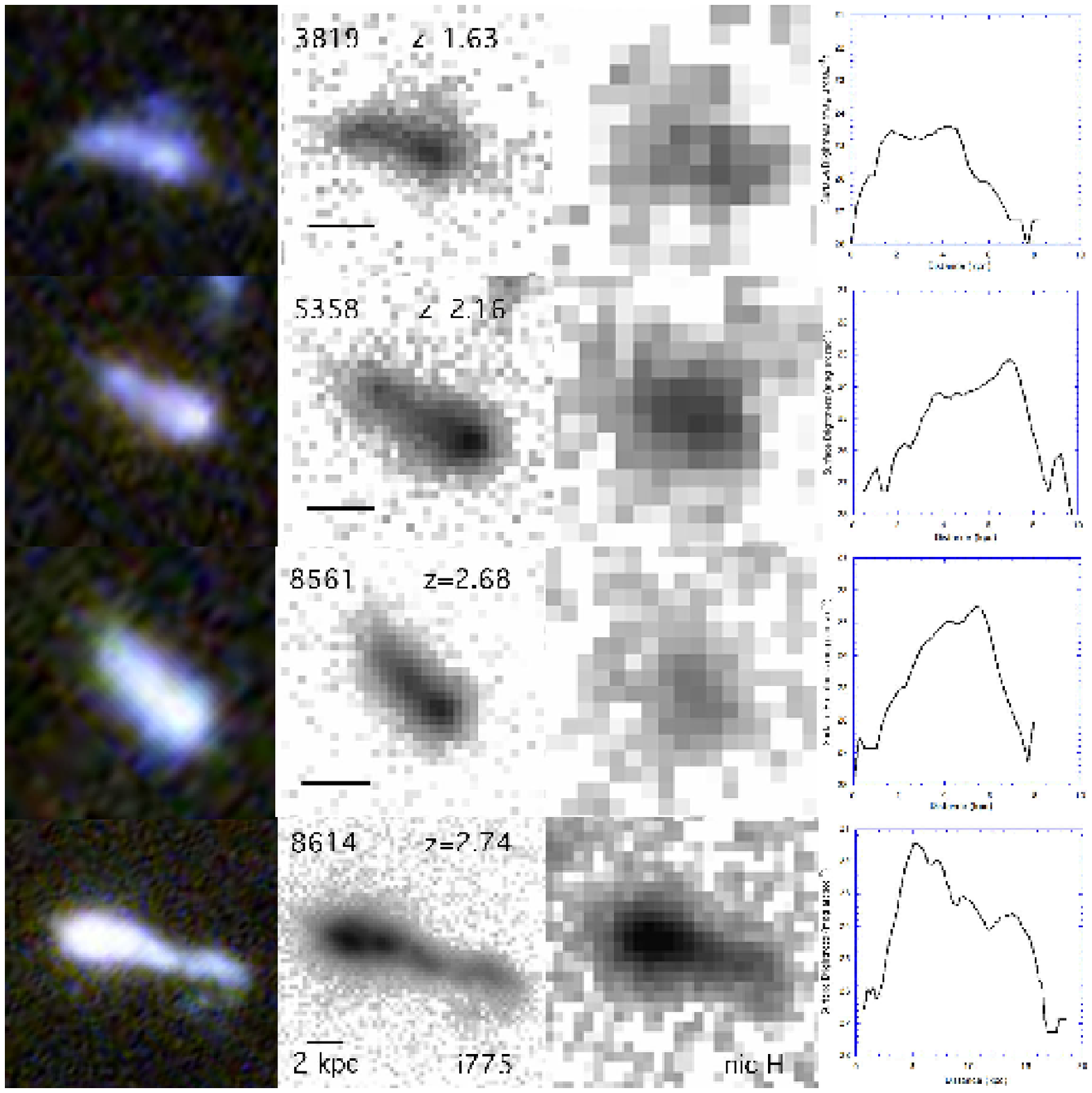} \caption{Four tadpole galaxies from the Hubble
UDF, with color Sky Walker images from the ACS filters on the left,
$i_{775}$ next, NICMOS H-band next, and an intensity scan through the
length of the galaxy in $i_{775}$ on the right.  UDF 8614 on the bottom
has a double-core head with the two cores aligned with the tail. [Image
quality degraded for arXiv]}\label{Fig1-tadpoles}\end{figure}

Figure \ref{Fig1-tadpoles} shows a collection of 8 tadpole galaxies at
ACS resolution in color, from the UDF Sky Walker\footnote{designed by
K. Jahnke and S.F. S\'anchez, AIP 2004} (on the left), in ACS $i_{775}$
(second from the left) and NICMOS $H$ (second from the right, with
pixels three times larger than in Sky Walker). On the right is an
intensity scan through the length of the galaxy as viewed by the ACS in
$i_{775}$ band. This class of objects has a bright head and a diffuse
tail. Many tadpoles have slightly different morphologies; some have
diffuse emission that is not particularly elongated to the side of a
clump. Sometimes the tail has a clump, or one could interpret the
structure as a double head. In all cases, there is a bright clump at
the end of a diffuse region.  Because photometric redshifts were
determined for whole tadpole galaxies and not the heads and tails
separately, we cannot tell if the different parts of a tadpole are
really separate galaxies. As a result, we interpret possible
differences in component redshift as differences in color, age, or
extinction.

Figure \ref{tadpoles_mag_vs_z} shows the redshift distribution of
apparent AB magnitude in $i_{775}$ band at full ACS resolution for the
tadpole heads and tails, in $i_{775}$ for the heads at the NICMOS
resolution, and in H-band for the heads from NICMOS. We reach a
limiting $i_{775}$ magnitude of $\sim29.5$.  Absolute magnitudes are
shown by the dotted lines, using distance moduli from a $\Lambda$CDM
cosmology \citep{spergel03}.  The NICMOS low-resolution measurements
are systematically brighter than the ACS high-resolution measurements
because of the inclusion of slightly more projected area at the lower
resolution. The tails are systematically brighter than the heads at the
same ACS resolution. Tails are often not observed well in NICMOS so we
do not discuss that measurement in this paper.

\section{Analysis}

\cite{bruz03} models of stellar population spectra were used to
determine redshifted model colors for comparison with the observed
colors of the heads and tails. A metallicity of 0.008 (equal to 0.4
solar) and the Chabrier IMF was assumed. We considered an exponentially
decaying star formation history with separate start times and decay
times for each clump and head.  The decay times considered were, in
Gyr, 0.01, 0.03, 0.1, 0.3, 1, 3, and 10. Intervening cosmological
hydrogen absorption \citep{madau95} is included, as well as internal
dust absorption using the wavelength dependence in \cite{calz00} with
the short-wavelength modification in \cite{leith02}. A $\Lambda$CDM
cosmology was assumed \citep{spergel03}. As mentioned above,
photometric redshifts come from \cite{rafel09}.

The rms differences between the model colors and the observed colors
were determined for each of a wide range of start times, decay times,
and extinctions. These rms differences were binned into groups with
values incremented by 0.1.  Among the groups with the lowest rms
differences, weighted average values of the start times (ages), decay
times, and extinctions were determined.  The weighting parameter is
$\exp(-0.5 \chi^2)$ where $\chi^2$ is the sum of the squares of the
color differences divided by the rms errors in the colors, as
determined from the pixel counts used for the magnitudes.  The mass
follows from the observed $i_{775}$ magnitude in comparison to the
model. The surface density of a tail is taken to be the tail mass
divided by the projected area of the measured region. For more details
of this method, see \cite{e09a,e09b}.

\section{Results}

\subsection{Masses and Ages}\label{sect:ma}

The masses of the tadpole heads and tails are shown as functions of $z$
in the left panel of Figure \ref{tadpoles_mass_vs_z}. For comparison,
clump masses in UDF clump-cluster and chain galaxies are shown in the
middle panel, and bulge masses in the clump clusters and chains that
have bulges are shown on the right \citep[from][]{e09a}. Bulges are
defined to be the reddest, and often the most luminous clumps; they are
usually, but not always, centralized. The tadpole head and
clump-cluster bulge masses were evaluated twice, once with J and H
NICMOS observations and 4-band ACS observations blurred to NICMOS
resolution (red circles), and again with only the 4-band ACS
observations at full resolution (blue dots). The small black dots in
the middle and right-hand panels are clumps and bulges, respectively,
in GOODS galaxies \citep{e09b}, which have 4 ACS bands and extend to
$z\sim1$.   There are relatively few GOODS measurements compared to UDF
measurements because clump clusters are extremely rare at the low
redshifts of the GOODS survey. The 4-band ACS measurements of tadpole
head mass with $0.03^{\prime\prime}$ pixel size (blue dots in the left
panel) are slightly smaller than the J-H NICMOS + ACS measurements with
$3\times$ larger pixels (red circles).  This is partly because the
larger pixels include more peripheral light, and partly because the
NICMOS IR bands include more old stars.

The average trends for $\log M$ (in $M_\odot$) versus $z$ on the left
in Figure \ref{tadpoles_mass_vs_z} are somewhat linear. We fit them to
the function $\log M=A+Bz$ with $A=5.86\pm0.50$ and $6.78\pm0.73$ for
ACS-only determinations of the head mass and ACS+NICMOS determinations
of the head mass, respectively. The slopes are $B=0.51\pm0.16$ and
$0.39\pm0.21$, respectively. Error bars for this linear fit are 95\%
confidence intervals. The general decrease in mass at low $z$ is a
selection effect related to size and cosmological surface brightness
dimming \citep{e09b}. For $z>1$, the average log of the tadpole head
mass calculated with ACS only is $7.69\pm0.69$. The average log mass
measured with ACS of 906 UDF clump cluster and chain clumps at $z>1$ in
the middle panel is $7.22\pm1.34$. The average log mass for 23 UDF
clump cluster and chain bulges at $z>1$ in the right-hand panel,
calculated with ACS only, is $8.11\pm0.43$. Thus the tadpole heads at
$z>1$ have an average mass that is larger than the average clump
cluster clump mass at the same redshift by a factor of $\sim3$, and
smaller than the average clump cluster bulge mass by a factor of
$\sim2.6$. The tadpole heads are not as massive as whole clump clusters
or chain galaxies in our surveys, which typically contain
$\sim10^{10}\;M_\odot$ or more in stars.

Tail masses in Figure \ref{tadpoles_mass_vs_z} were determined from
rectangular areas enclosing the whole tail, including some apparent sky
regions but not the head, to the extent that excluding the head was
possible.   A linear fit to the log of the tail mass in Figure
\ref{tadpoles_mass_vs_z} gives $\log M = (6.93\pm0.64)+(0.36\pm0.20)z$.
The sum of the head and tail for individual galaxies scales with
redshift as $\log M_{\rm sum}=(6.82 \pm0.54)+(0.48\pm0.17)z$. At $z=2$,
this summed mass is $6\times10^7\;M_\odot$. The average of the log of
the ratio of the tail to the head mass for individual galaxies is
$0.67\pm0.86$, meaning that tails are more massive than heads by a
factor of $\sim4.7$.

Root mean square deviations in the logarithm of the mass are shown in
Figure \ref{tadpoles_rms} on the left, averaged over bins of redshift
for all of the objects. The three panels are for different types of
measurements: ACS measurements of the heads (top), ACS+NICMOS
measurements of the heads (middle), and ACS measurements of the tail
(bottom). The rms values of $\log M$ represent deviations among the
mass results for all of the models within the lowest bins of rms
deviations in the color, as discussed above.  The rms values are lowest
for the ACS+NICMOS measurements, and there is a slight increase in rms
with redshift.

Figure \ref{tadpoles_age_vs_z} shows the tadpole head ages in
comparison to the ages of clumps and bulges in clump cluster and chain
galaxies. The green dashed lines represent the age of the universe as a
function of redshift.  Derived ages are less certain than masses
because of an ambiguity between reddening from age and reddening from
extinction (these two effects compensate for each other in the case of
mass).  Root mean square values for the logarithm of the age are shown
on the right in Figure \ref{tadpoles_rms}; they are larger than the
$\log M$ rms values by about 50\%. The ages of the tadpole heads span
the same large range as the ages of the bulges in clump clusters and
chain galaxies. The heads are older than the bulk of the clumps at low
redshift, but comparable to the ages of clumps at $z\sim2-4$, which are
also comparable to the ages of bulges there.

Figure \ref{tadpoles_sb_vs_z} shows the projected mass surface
densities of tadpole tails versus redshift (black dots), compared with
the mass surface densities of the interclump media in four types of
galaxies in the GOODS fields, from \cite{e09b}. These four types are
labeled in the panels. Two-arm spirals and flocculent spirals resemble
spiral galaxies today; clump clusters are composed primarily of several
massive clumps of star formation with little interclump medium, and
clump clusters with red disks have the same clumpy star formation but
there is a red old-star component between them. We suggested in
\cite{e09b} that there is an evolutionary sequence from clump clusters
with no evident interclump medium to clump clusters with red interclump
media, to spirals, on the basis of the mass surface density and age of
the interclump regions, in addition to the presence of bulges in spiral
galaxies, which seems to be a later phase than a clump cluster
\citep{e09a}.  The GOODS galaxies extend to $z\sim1$ while the UDF
galaxies extend further in redshift. Both have an increasing trend of
surface density with redshift from selection effects related to
cosmological surface brightness dimming.  The rms errors in the fits to
the surface density are the same as the rms errors in the fits to the
mass.

The mass surface densities in tadpole tails were determined from
rectangular regions entirely enclosed in the tails, so they represent
the values in the tail centers. Figure \ref{tadpoles_sb_vs_z} indicates
that the mass surface densities in tadpole tails are generally less
than the mass surface densities in spiral and flocculent galaxies, by a
factor of $\sim10$. Thus the tails are not normal galaxy disks (but
they could be low surface-brightness disks). The tadpole tails are also
slightly lower in surface density than the red parts of clump clusters
with red interclump regions, and also lower than the interclump regions
between the clumps of pure clump clusters. These trends are evident
only in the small region of overlap in the figure, which is at low
redshift.  We do not have similar measurements for interclump regions
in high redshift clump clusters from the UDF because these regions are
generally very faint. We see the tadpole tails in the UDF because they
are isolated from bright clumps.

Figure \ref{tadpoles_ageinterclump_vs_z} shows the ages of the tadpole
tails compared with the ages of the interclump regions of the 4 types
of galaxies in our GOODS study.  These region are compared because they
are all somewhat diffuse and outside the obvious star formation clumps.
The figure indicates that all of the ages are about the same, in the
range from 0.01 to 1 Gyr, with a concentration around 0.1 Gyr for the
high-redshift tadpole tails.

A histogram of the difference in the log of the age between the head
and the tail for individual galaxies ($\log{\rm head\;age}-\log{\rm
tail\;age}$) is shown in Figure \ref{tadpoles_agedif}.  The average
difference is $-0.3\pm0.9$, suggesting slightly younger heads than
tails, but this difference is essentially zero within the errors.

The head and tail densities can be estimated from the masses, column
densities, and sizes. The average size of a tadpole head is measured to
be $\sim0.3\pm0.4$ kpc, corrected for the ACS point spread function
(Section \ref{comp}). For a typical mass of $10^8\;M_\odot$ (Figure
\ref{tadpoles_mass_vs_z}), the head has a density of luminous stars
$\sim3.7\;M_\odot$ pc$^{-3}$, and a column density $\sim1100\;M_\odot$
pc$^{-2}$. We can do this more accurately considering each galaxy
separately and counting only those with resolved heads. Then the
average head density is $2.0\pm4.4\;M_\odot$ pc$^{-3}$, and the average
head column density is $880\pm1800\;M_\odot$ pc$^{-2}$. The transverse
size of a tail is about the same as the head size. The average tail
surface density is $69\pm83\;M_\odot$ pc$^{-2}$ in Figure
\ref{tadpoles_sb_vs_z}, so the average tail density is this surface
density divided by the typical size, or $0.2\;M_\odot$ pc$^{-3}$. If
the tail density is determined for each galaxy separately, then the
average is $0.12\pm0.13\;M_\odot$ pc$^{-3}$, a factor of 16 less than
the head density.

Histograms of the extinctions in the heads and tails are shown in
Figure \ref{tadpoles_hisext}. These were obtained from the stellar
population fits to the 4 ACS passbands, along with the masses, ages,
and star formation decay times (not shown). The average extinctions for
the heads and tails are $1.7\pm1.3$ mag. and $1.7\pm1.2$ mag.,
respectively. Extinctions and ages are the most uncertain parts of the
model fits.  The ACS+NICMOS fits to the head regions also gave
extinctions, but because NICMOS could not detect most of the tails and
was not used for the tail model fits, this figure shows only the ACS
extinctions in comparing the heads and the tails.

\subsection{Companions}\label{comp}

We studied the environments of tadpoles by counting companions within a
fixed rest-frame projected separation and a fixed redshift interval.
The companions were UDF galaxies larger than 100 square pixels in area,
since the tadpoles themselves are larger than 10 pixels in length. We
considered only the UDF galaxies with photometric redshifts in the
latest compilation based on supplemental uv data \citep{rafel09}. This
is consistent with our use of the same redshift catalog for the
tadpoles themselves.

Figure \ref{tadpoles_udfclos_onlymarc_100kpc_z0.2} (left) shows, in
red, a histogram of the number of companions to tadpole galaxies within
a projected distance of 100 kpc at the distance of the tadpole and
within a redshift interval of 0.2. The right hand axis is used. It also
shows, in blue, a histogram of the number of companions for all
galaxies in the same UDF catalog used for the companions, i.e., larger
than 100 px$^2$ area with redshifts in \cite{rafel09}. The left-hand
axis is used for this. The histograms are very similar, suggesting that
the number distribution of companions around tadpole galaxies is about
the same as the number distribution of companions around any other
large galaxy. This would imply that tadpoles do not have an excess or
lack of companions compared to other galaxies. The right-hand panel
shows the same neighbor counts in a cumulative distribution, from which
a Kolmogorov-Smirnov test was performed.  We find a 74\% probability
that the tadpole companions and the field-galaxy companions are drawn
from the same near-neighbor distributions. Other separations and
redshift intervals give the same result: for separations within 100 kpc
and a redshift interval of 0.8, the KS probability is 58\%; for 200 kpc
and $\Delta z=0.2$, 53\%, and for 200 kpc and $\Delta z=0.8$, 64\%.
Also for these other limits, the peaks in the near-neighbor histograms
shift toward more neighbors, as expected for the bigger space volumes
considered. For the 4 cases, respectively, the peaks of the
distributions are centered at 1, 6, 7, and 26 companions.  In a more
extreme case with a separation of 50 kpc and a redshift interval of
0.2, there are very few galaxies in the neighborhood: the peak in the
distribution of the number of companions is at 0 companions and the KS
probability that the tadpoles are drawn from the same distribution as
the field galaxies is nearly 100\%.

To further test the similarity of tadpole neighborhoods with those of
field galaxies, we added fake galaxies around the tadpoles with certain
probabilities to see how this affects the KS tests. Recall that for 100
kpc distance and a redshift interval of 0.2, the KS probability that
the real tadpole neighbors and the field neighbors are from the same
distribution is 74.3\%. If we add one extra companion to 2\% of the
tadpoles, then this probability drops to 26.6\%. For an extra companion
added to 5\% of the tadpoles, it drops to 9.3\%, for 10\% of the
tadpoles, 0.084\%, and for 50\% of the tadpoles, 0.0093\%. Thus we
cannot tolerate an additional companion around even 5\% of the tadpoles
before the companion distribution begins to look significantly
different from the field galaxy companion distribution. The reason for
this is that additional companions shift the tadpole histograms in
Figure \ref{tadpoles_udfclos_onlymarc_100kpc_z0.2} slightly to the
right relative to the distribution for all galaxies. The histograms are
so sharply peaked, however, than even a slight shift for the tadpoles
causes the histogram to significantly exceed the field galaxy histogram
at high numbers.

We conclude from these tests that tadpole galaxies have normal neighbor
distributions; they are not significantly close to companions larger
than 10 pixels in size. They could still be mergers, however, with the
merged galaxies unresolved in the head for the majority of cases where
only one head clump is observed.

The case of mergers is investigated next. Figure
\ref{tadpoles_hisheadsize} shows the size distributions of the tadpole
heads and tails.   The head size is defined to be the square root of
the difference between the area in pixels and the area of the FWHM of
the point spread profile. The FWHM point spread profile was measured
from stars at V$_{606}$ to be 3.08 pixels. The corrected angular size
is converted into kpc using the redshift. The average head size
(including those with effectively zero size because they are at the
resolution limit) is $0.31\pm0.36$ kpc. This is relatively small for a
merger; each tadpole would have to have its merging sub-galaxy close to
perigalacticon. The average tail length is $3.9\pm1.7$ kpc, also small
for a tidal tail by local standards.

We also looked for double-core or clumpy heads. From the list of
tadpoles in \cite{e05}, 13\% have something that might be called a
double-core head, 3.3\% have multi-core (lumpy) heads, and 6.5\% have
disky (elongated) heads. Figure \ref{3_wiggly_tads} shows examples of
tadpoles with clumpy heads or multiple-clump structure throughout.
These multi-core cases could be interactions or mergers, or they could
just be clumpy, single-galaxy heads. UDF 9543 has a wiggly tail, which
is rare. The tadpole UDF 8614 in Figure 1 has a double-core head and
wiggly tail too (when viewed with the right contrast, the two cores in
the head are aligned with the tail). The wiggly tails in these two
cases could be the result of variable external pressure forces or
orbital motions inside the heads, which gravitationally drag the tails
around with them. Most tadpoles in our present survey do not have
double cores in their heads. Perhaps higher resolution observations
will show more double cores.

Another consideration is the orientation of the tadpole tails relative
to nearby galaxies. Figure \ref{tadpoles_angle_vs_separation} shows the
distribution of angles, measured at the midpoint of the tail, between
the midpoint of the head and the midpoint of a companion galaxy, versus
the distance between the head and the companion normalized to 100 kpc
at the distance of the tadpole.  Companions are considered within a
redshift interval $\Delta z=\pm0.2$. When this angle is $0^\circ$, the
tadpole head points toward the companion, and when it is $180^\circ$,
the head points away from the companion. These are projected angles, so
the numbers of tadpole-companion combinations in each $45^\circ$
quadrant of this plot would be about equal for a random distribution of
orientations. This appears to be the case, suggesting that tadpoles are
not pointing in any particular direction relative to their companions,
regardless of the companion distance. We found the same random
orientations for various combinations of maximum projected separation
and maximum redshift interval (i.e., separations of 100 and 200 kpc and
maximum $\Delta z$ of 0.2 and 0.8).

Tadpole tails can be curved, flared, or clumpy. In 92 tadpoles from our
UDF sample, 22\% are flared and about half of these are also clumpy in
the flared regions, 13\% of the tails are curved and most of these have
no significant clumps, and 21\% are straight and clumpy. The rest
(42\%) are straight without significant clumps.  Curved tails have only
minor curvature, around $30^\circ$ at most.

Companions to high redshift galaxies were also studied by
\cite{conselice09}, who looked at ACS UDF dropout paired galaxies in
B$_{450}$, V$_{660}$, and i$_{775}$ bands with magnitudes
$z_{850}<28.5$.  These samples correspond to redshifts equal to
approximately 4 (320 galaxies), 5 (137 galaxies), and 6 (126 galaxies).
They defined close pairs as two galaxies in the same dropout category
with projected separations less than 20 kpc (for $H_0=75$ km s$^{-1}$
kpc$^{-1}$). The dropout pair fractions for the three passbands are
$0.21\pm0.03$, $0.19\pm0.04$ and $0.16\pm0.05$, respectively. They also
studied the asymmetry index for a subsample of these galaxies with
$z_{850}<27.5$ (69, 43, and 21 galaxies, respectively) and found
similar fractions with an asymmetry indicative of a merger:
$0.23\pm0.05$, $0.19\pm0.05$, and $0.19\pm0.13$, respectively. In the
present study, we have excluded galaxy pairs in our selection of
tadpole shapes (``doubles'' were a different morphology class in our
UDF catalog).  We also exclude galaxies with peculiarities or
asymmetries that are not like tadpole shapes; tadpoles are well defined
and usually symmetric around one axis. Thus the comparison with the
work of Conselice \& Arnold is not straightforward. Overall, the
tadpole fraction in the UDF (10\% in our catalog) is about half of the
pair or asymmetry fractions found by Conselice \& Arnold. Our search
for companions within 50 kpc, 100 kpc, or 200 kpc and a small range in
redshifts covers a much larger volume than the paired companions
searched by Conselice \& Arnold, which are within 20 kpc. Tadpoles are
not, by definition, members of pairs or such strong mergers that they
become highly distorted. Because there is also no evidence for an
excess or lack of companions at 50 kpc or beyond, the tadpoles do not
seem to get their main structure from interactions.

Although we see no evidence for interactions, mergers, or nearby
companions in most tadpole galaxies, we cannot rule out interactions
and mergers as a cause for their structure. Interactions could involve
objects smaller than 100 pixels in area or with uncertain redshifts,
which were excluded from our neighbor list. They could all be
late-stage mergers with unresolved double cores, although the tadpole
fraction of 10\% would require a high and continuous merger rate, as
mentioned in the introduction.

\subsection{Internal Dynamics}\label{sect:dyn}

Understanding the nature of tadpole galaxies requires dynamical
information.  \cite{forster09} include 4 tadpoles in their SINFONI
spectroscopic survey of high-redshift galaxies. These galaxies are
SSA22a-MD41 and Q2343-BX389, shown in their Figure 24, and Q2346-BX405,
and Q2346-BX482, shown in their Figure 25.  In order, their
spectroscopic redshifts are 2.17, 2.17, 2.03, and 2.26, which are in
the same range as our tadpoles. Their total stellar masses from SED
fits are $0.72\times10^{10}\;M_\odot$, $4.40\times10^{10}\;M_\odot$,
$1.58\times10^{10}\;M_\odot$, and $1.69\times10^{10}\;M_\odot$, which
are larger than the masses of our tadpole heads by nearly 2 orders of
magnitude (see Figure \ref{tadpoles_mass_vs_z}).  Their dynamical
masses from gas emission lines are, respectively,
$6.9\times10^{10}\;M_\odot$, $14\times10^{10}\;M_\odot$,
$2.8\times10^{10}\;M_\odot$, and $13\times10^{10}\;M_\odot$, which are
larger than their SED stellar masses by factors of 9.6, 3.2, 1.8, and
7.7.  These are reasonable factors for dark matter halos.

\begin{figure}\epsscale{1.}
\plotone{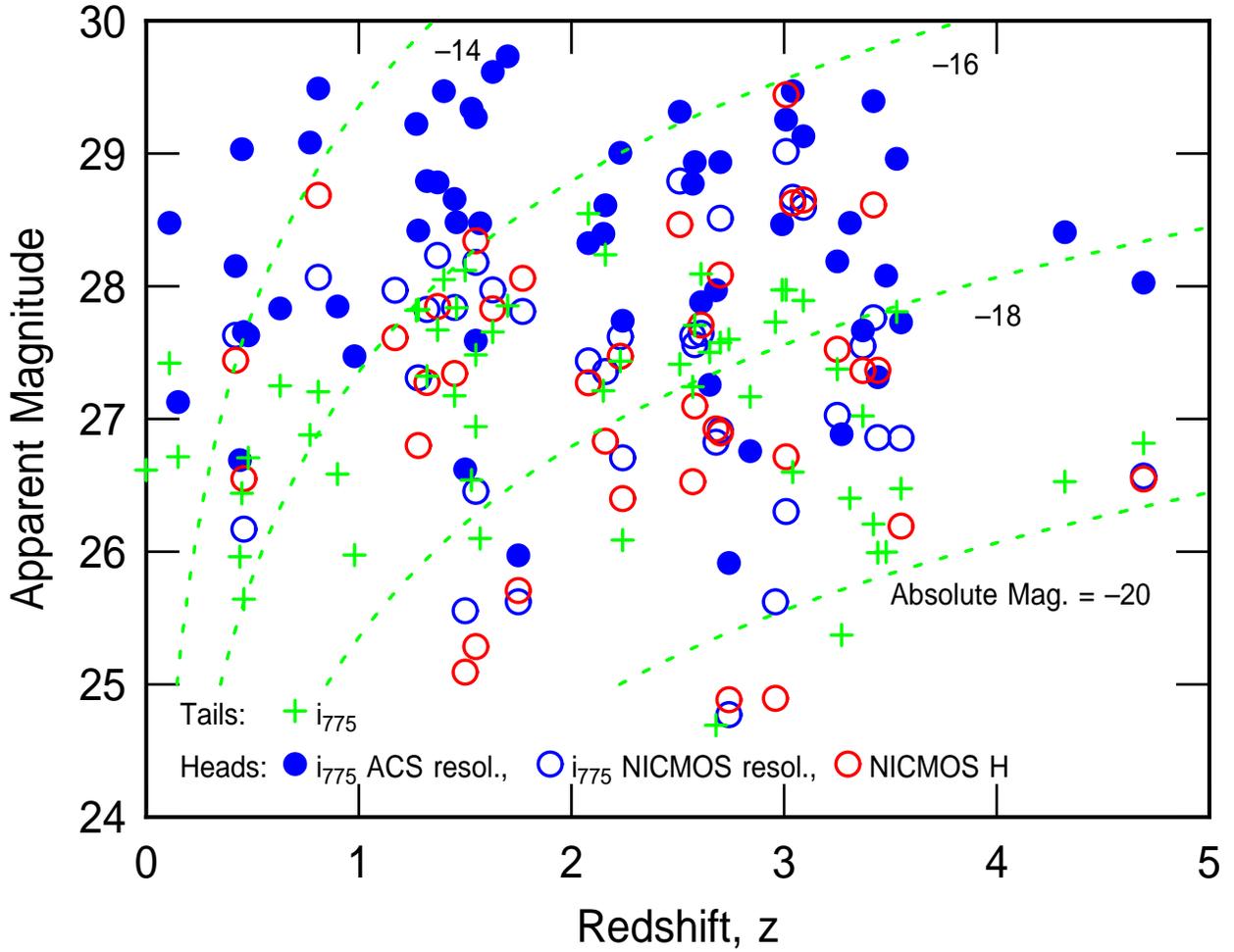} \caption{Magnitudes of the heads (dots) and tails
(plus symbols) at ACS resolution in $i_{775}$ band, $i_{775}$ band
magnitudes of the heads at the NICMOS resolution (blue circles), and
NICMOS H-band magnitudes of the heads (red circles). The dotted lines
show absolute magnitudes, as
indicated.}\label{tadpoles_mag_vs_z}\end{figure}

\begin{figure}\epsscale{1}
\plotone{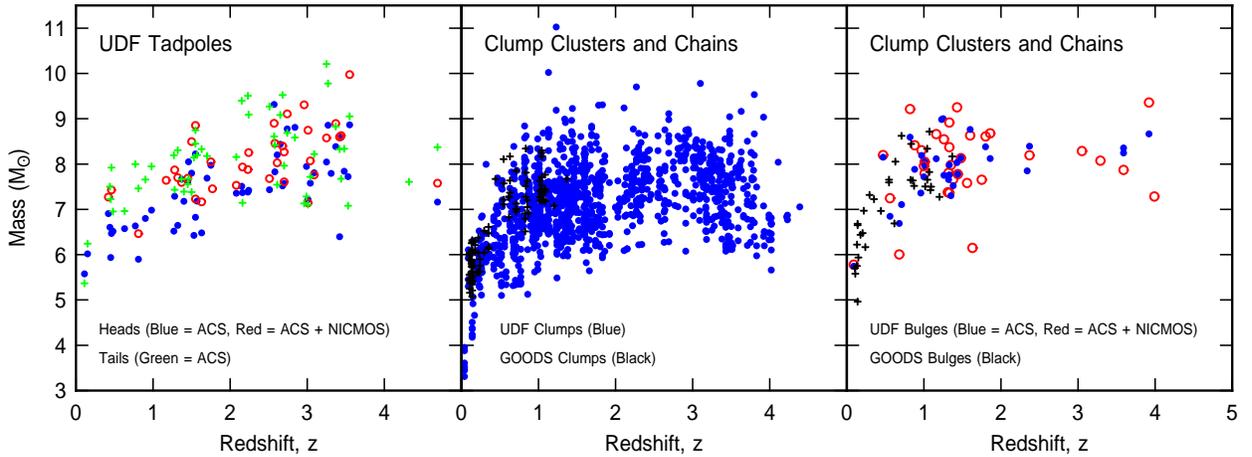} \caption{(left) Masses of the tadpole heads and tails
versus redshift. Tail masses (green plus symbols) used only the full
resolution ACS bands. Head masses were evaluated twice, once with
ACS-only filters at full resolution (blue dots), and again with NICMOS
filters combined with ACS filters blurred to NICMOS resolution (red
circles). (middle) Masses of clumps in clump cluster and chains
galaxies in the UDF (blue dots) and GOODS fields (black). (right)
Masses of bulges or bulge-like objects in clump clusters and chains in
the UDF (blue dots with ACS resolution; red circles with NICMOS filters
and ACS filters at NICMOS resolution), and GOODS (black).  The clump
cluster and chain measurements are from
\cite{e09a,e09b}.}\label{tadpoles_mass_vs_z}\end{figure}

\begin{figure}\epsscale{1}
\plotone{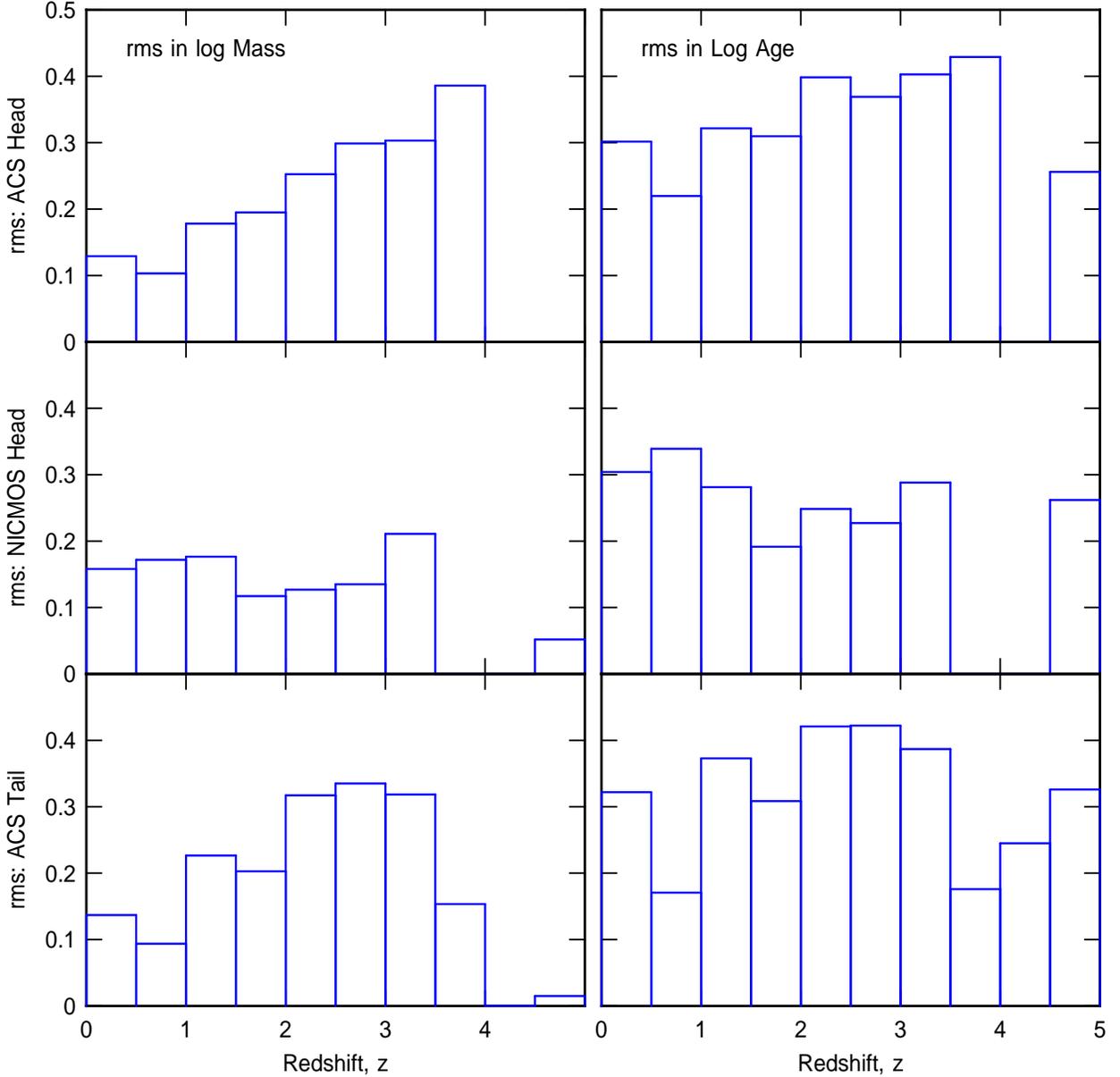} \caption{(left) The rms deviations in the log of the
measured masses are shown, averaged over bins of 0.5 in redshift. The
top, middle and bottom panels are for ACS measurements of the heads,
ACS+NICMOS measurement of the heads, and ACS measurements of the tails.
The masses are calculated in $M_\odot$. For the ACS+NICMOS
measurements, the $\log M$ of the heads is accurate to about $\pm0.15$,
which corresponds to a factor of 1.4. (right) The rms deviations in the
log of the ages for the same three measurement cases. Age rms values
are higher than mass rms values because reddening effects from age and
extinction partially cancel in the determination of mass.}
\label{tadpoles_rms}\end{figure}

\begin{figure}\epsscale{1}
\plotone{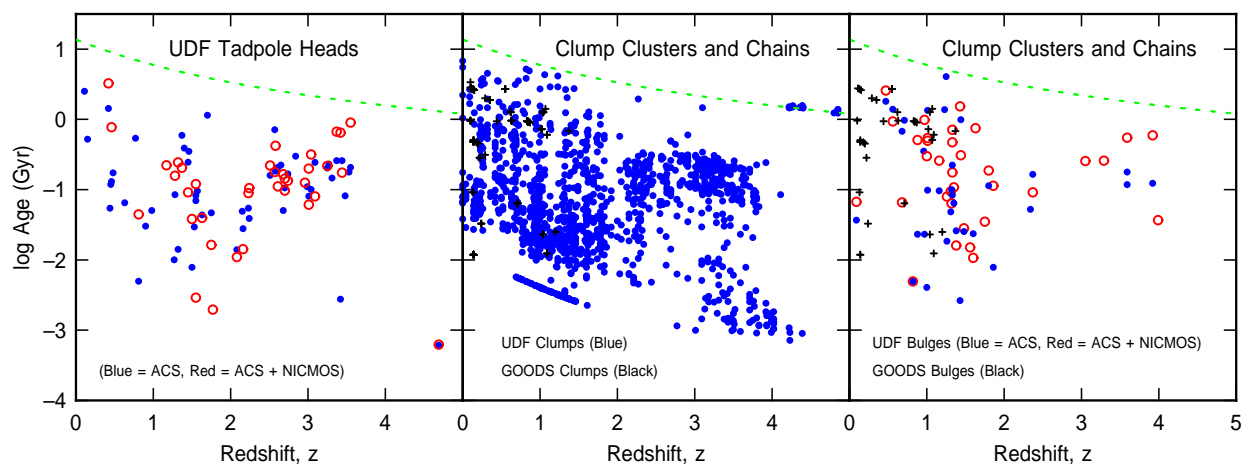} \caption{(left) Ages of the tadpole heads versus
redshift. ACS-only filters are blue dots, and NICMOS filters combined
with ACS filters blurred to NICMOS resolution are red circles. (middle)
Ages of clumps in clump cluster and chains galaxies from the UDF (blue
dots) and GOODS fields (black). (right) Ages of bulges or bulge-like
objects in clump clusters and chains in the UDF (blue dots with ACS
resolution; red circles with NICMOS filters and ACS filters at NICMOS
resolution), and GOODS (black).  The clump cluster and chain
measurements are from \cite{e09a,e09b}. The green dashed lines
represent the age of the universe as a function of
redshift.}\label{tadpoles_age_vs_z}\end{figure}

\begin{figure}\epsscale{1}
\plotone{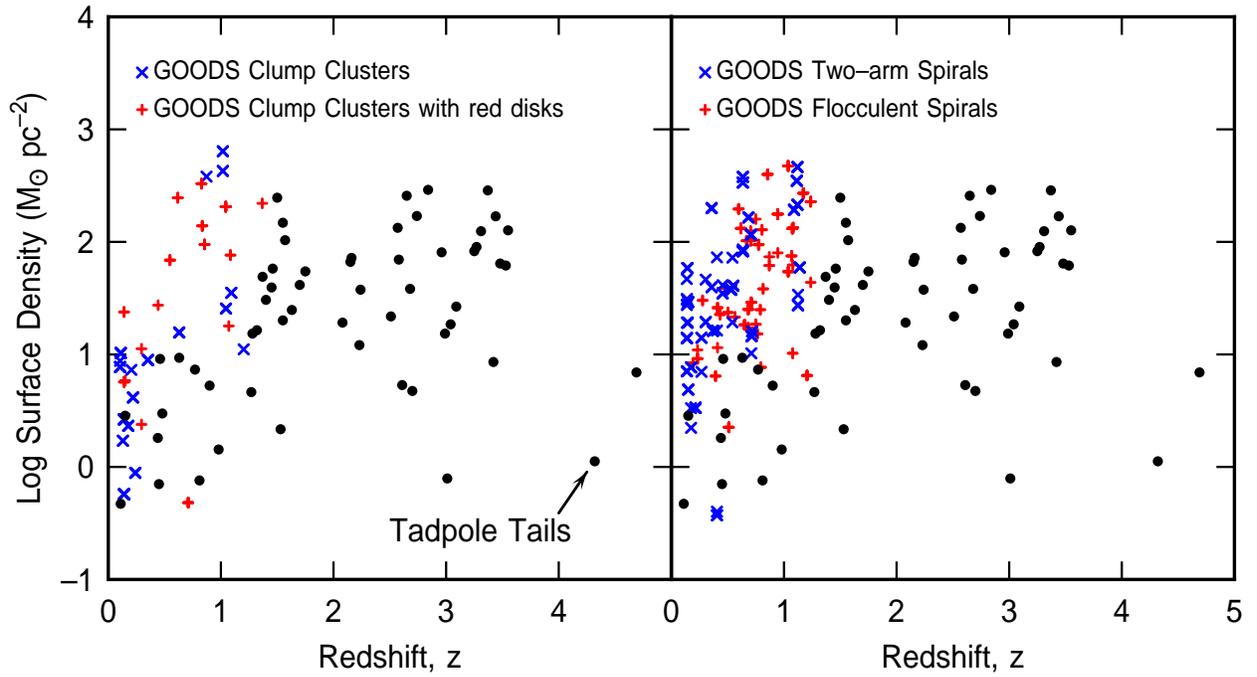} \caption{Mass surface densities of the tadpole tails
(black dots, repeated in both panels) compared, on the left, to the
interclump surface densities in GOODS clump clusters without red
underlying disks (blue x-symbols) and with red underlying disks (red
plus symbols). On the right, the tadpole tail surface densities are
compared to the interclump surface densities in GOODS flocculent and
grand design spirals.}\label{tadpoles_sb_vs_z}\end{figure}

\begin{figure}\epsscale{1}
\plotone{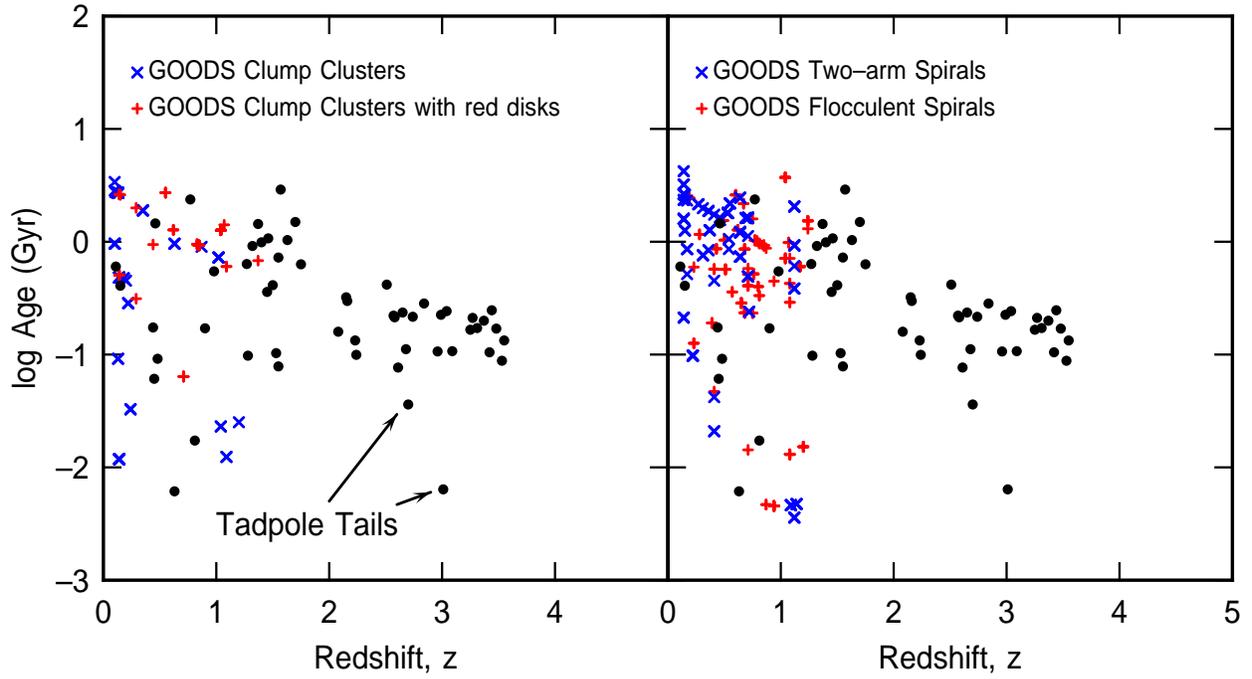} \caption{Ages of the tadpole tails (black dots,
repeated in both panels) compared, on the left, to the interclump ages
in GOODS clump clusters without red underlying disks (blue x-symbols)
and with red underlying disks (red plus symbols). On the right, the
tadpole tail ages are compared to the interclump ages in GOODS
flocculent and grand design
spirals.}\label{tadpoles_ageinterclump_vs_z}\end{figure}

\begin{figure}\epsscale{1}
\plotone{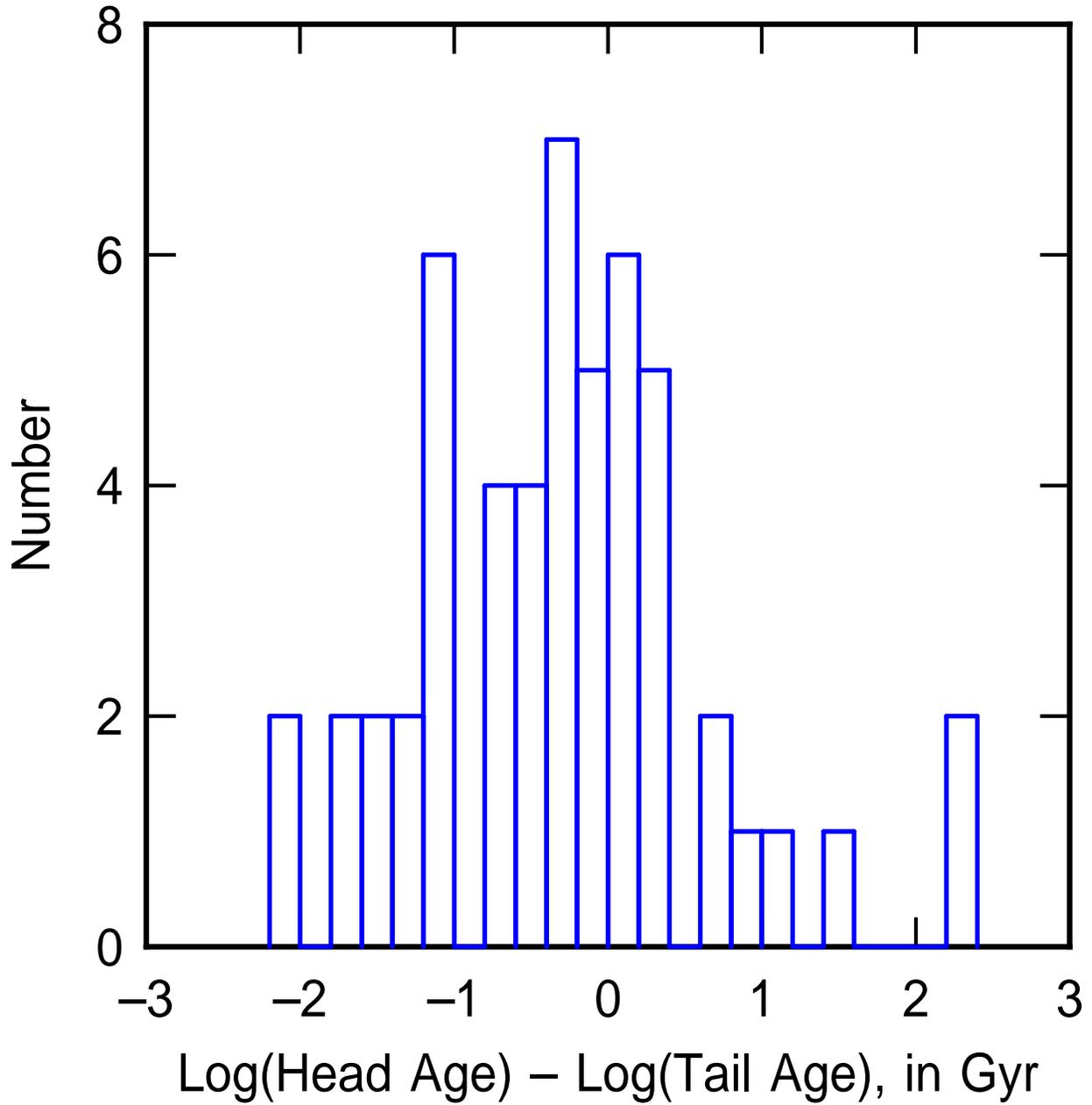} \caption{The distribution of the differences between
the logs of the head and tail ages for individual tadpole galaxies. The
average is about zero, meaning that the two components have about the
same age on average.}\label{tadpoles_agedif}\end{figure}

\begin{figure}\epsscale{1}
\plotone{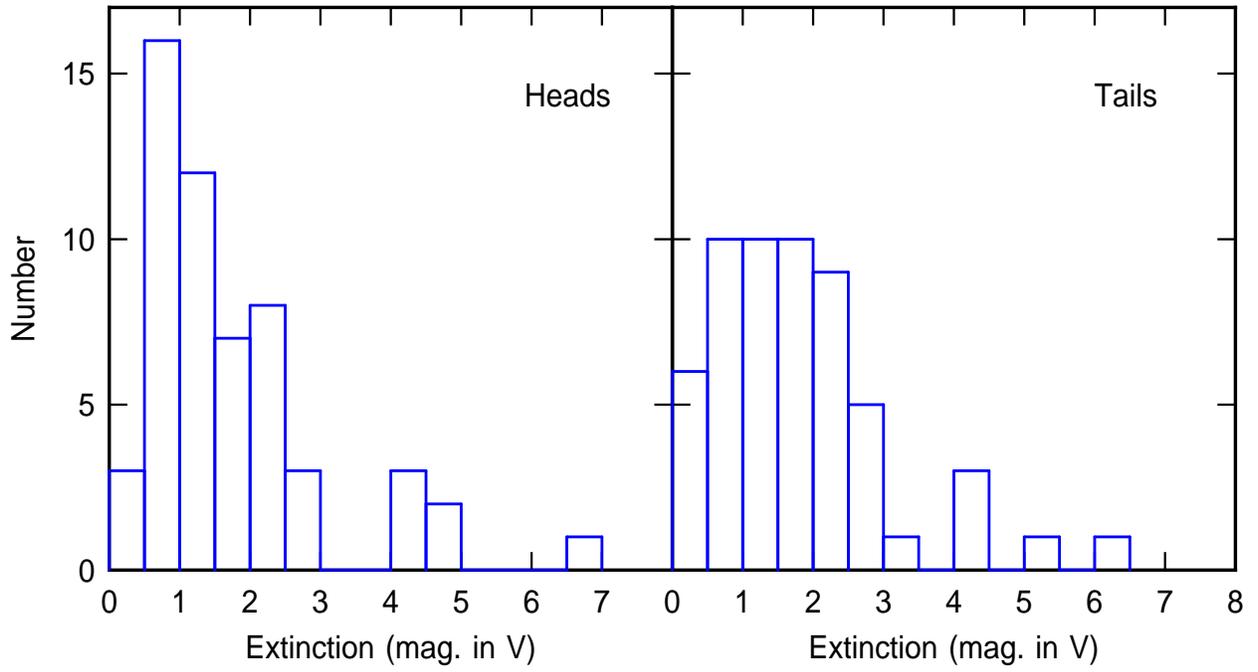} \caption{The distributions of the extinctions in
V-band for the tadpole heads and tails, obtained as part of the fits to
mass and age.}\label{tadpoles_hisext}\end{figure}

\begin{figure}\epsscale{1}
\plotone{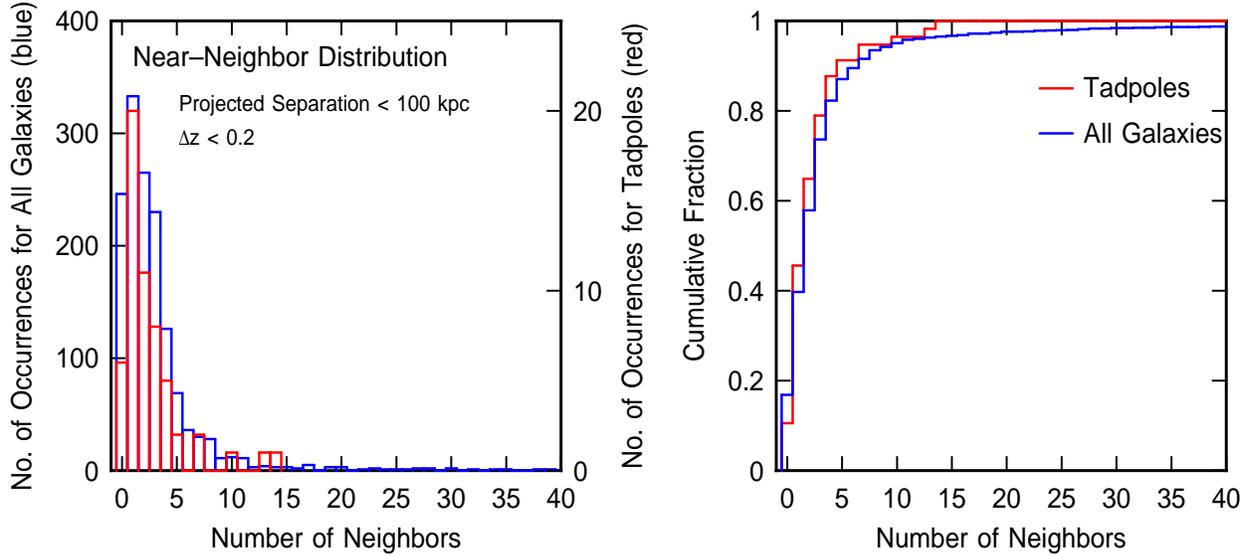} \caption{(left) The distribution function of
companion galaxy counts within 100 kpc projected distance from the
tadpoles (red histogram, right-hand axis), and within a redshift
interval of $\Delta z=\pm0.2$. A companion is defined to be a galaxy in
the UDF with a redshift in \cite{rafel09} and an area larger than 100
px$^2$. The left-hand axis (blue histogram) shows the same companion
distribution with respect to all galaxies larger than 100 px$^2$ in the
same redshift catalog.  The two distributions are about the same.
(right) Normalized cumulative distributions obtained by integrating the
histograms on the left. This figure, and a Kolmogorov-Smirnov test
based on this figure, suggest that tadpole galaxies do not have an
excess or deficit of neighbor galaxies with similar or larger sizes.
}\label{tadpoles_udfclos_onlymarc_100kpc_z0.2}\end{figure}

\begin{figure}\epsscale{1}
\plotone{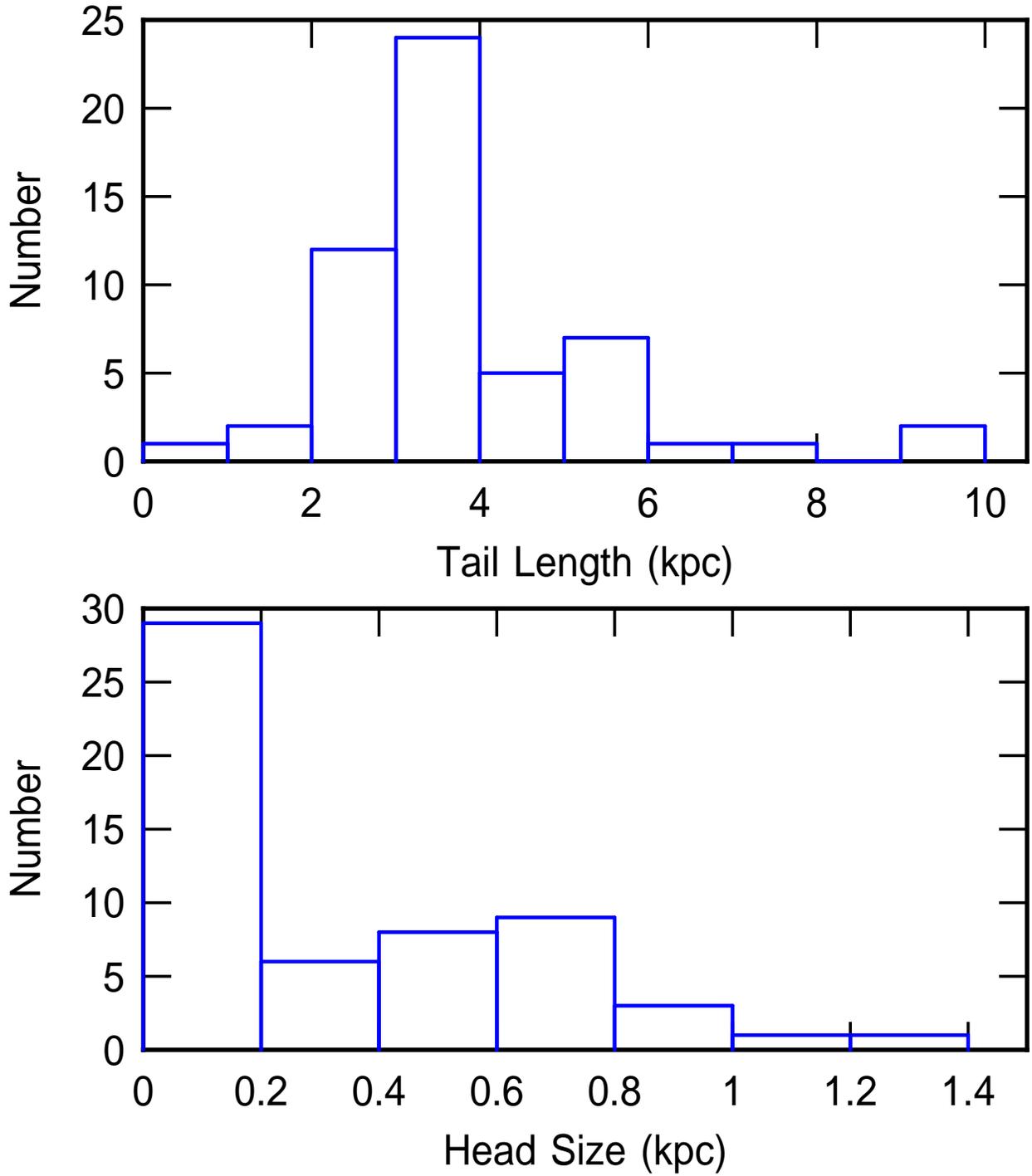} \caption{The distribution of tadpole head and tail
sizes, measured as the square roots of the areas, considering the
conversion from angular size to physical size at the redshift of the
galaxy and correcting for point spread
function.}\label{tadpoles_hisheadsize}\end{figure}

\begin{figure}\epsscale{1}
\plotone{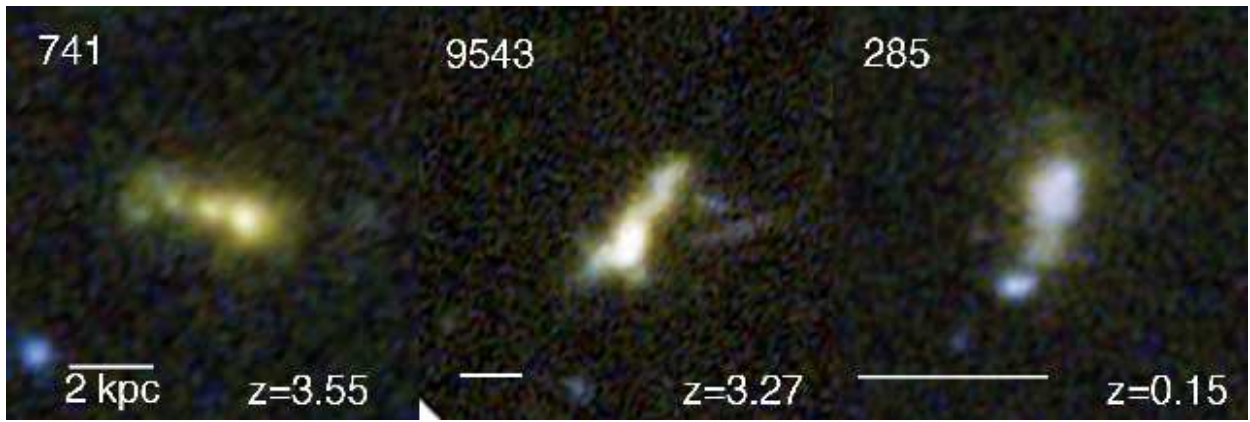} \caption{Three examples of tadpoles with clumpy
structure that might indicate interactions between the clumps. Most
tadpoles show single-core heads, suggesting they are not early-stage
mergers. [Image quality degraded for
arXiv]}\label{3_wiggly_tads}\end{figure}

\begin{figure}\epsscale{1}
\plotone{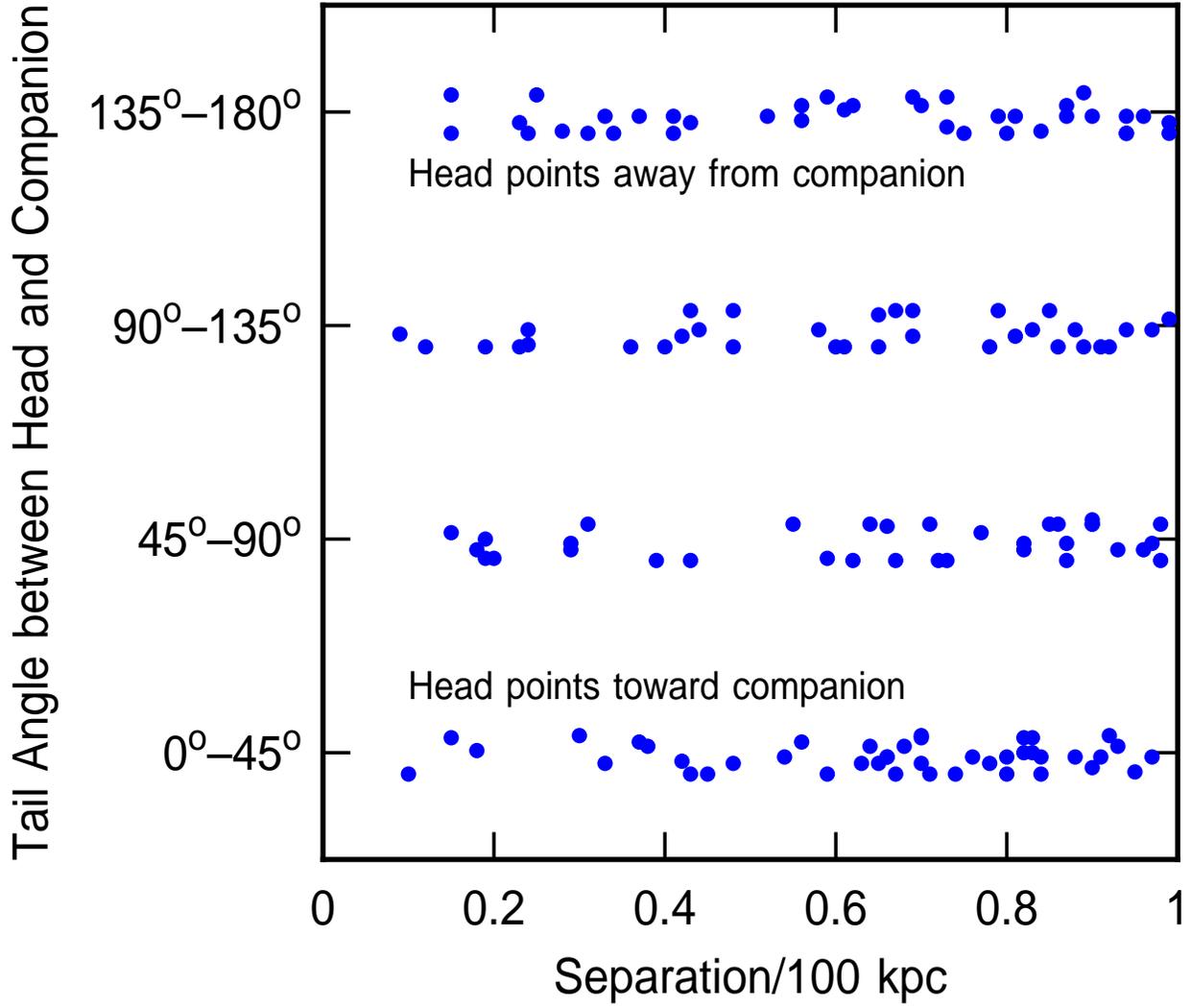} \caption{Distribution of the angle measured at the
tail, between the head and a companion, for all companions within 100
kpc projected separation and within a redshift interval of $\Delta
z=0.2$. Each dot is a different combination of tadpole and companion
galaxy.  The convention for angle is that $0-45^\circ$ means the head
is pointing toward the
companion.}\label{tadpoles_angle_vs_separation}\end{figure}

The higher masses in the F\"orster Schreiber et al. study reflect the
brighter magnitude limits of their surveys. UDF tadpoles are small by
comparison. The NICMOS H$_{\rm AB}$ magnitudes of the heads of our
tadpole galaxies range from 25 to 29 (see Figure
\ref{tadpoles_mag_vs_z}). The H$_{\rm Vega}$ magnitudes of the first 2
tadpole galaxies from the F\"orster Schreiber et al. list above are
$21.27\pm0.05$ mag and $21.75\pm0.10$ mag, and for the last galaxy,
$20.98\pm0.07$ mag. Adding the magnitude correction of 1.31
\citep{stan05} to convert from Vega to AB magnitudes, these become
$22.58$ mag, $23.06$ mag, and $22.29$ mag, respectively. The average of
these is $4.3$ mag brighter than the middle range for our H-band
magnitudes of tadpoles, and this difference explains the factor of
$\sim100$ larger stellar masses in the F\"orster Schreiber et al.
tadpoles. Evidently, tadpole structures are not limited to low-mass
galaxies, although the UDF tadpoles tend to be low-mass.

The position-velocity diagrams from gas emission lines for SSA22a-MD41,
Q2346-BX405, and Q2346-BX482 in F\"orster Schreiber et al. are fairly
straight throughout both the head and the tail parts, suggesting solid
body rotation or some other uniform gradient in the line-of-sight
motion. Q2343-BX389 has a 2-component position-velocity distribution,
with a straight part in the head and another straight part with a lower
gradient in the tail. The two velocities in Q2343-BX389 join well in
the middle, where the head meets the tail, so the components are not
likely to be from two different galaxies on the same line of sight.
Looking closer, two of the tadpoles with the nearly straight
position-velocity distributions, SSA22a-MD41 and Q2346-BX482, also have
slight kinks where their tails meet their heads, although within each
component, the position-velocity distributions are straight.  Only
Q2346-BX405 has a position-velocity orientation that is the same for
the head and the tail. The 3 cases with velocity kinks could be
signatures of complex dynamics involving galaxy interactions. but they
could also arise from other forcings.

The model for tadpoles implicitly assumed by Foster Schreiber et al. is
that each system is a single rotating disk and the dynamics reflects
the potential well from a common dark matter halo.  There is some
preference for this model. The dynamical mass is proportional to the
square of the velocity extent multiplied by the first power of the
spatial extent. For a nearly straight position-velocity distribution,
taking only half of the extent by cutting out the tail leads to a
dynamical mass for the head that is smaller than the total by $1/8$.
This would mean that the dynamical masses in the head regions would be
comparable to or less than the total stellar masses given above (recall
that we determined the total-to-stellar mass ratios above, and the
average is 5.5). Because the heads represent a significant fraction of
the total luminosities (e.g., for our tadpoles, the heads represent
30\%--50\% of the total luminosity, from Figure
\ref{tadpoles_mag_vs_z}), it is unlikely that the heads alone trace the
stellar+dark matter masses over half the extent of the
position-velocity distribution, and the tails trace an unbound tidal or
stripped feature unconnected with dark matter. Only for Q2343-BX389,
which has the strongly kinked position-velocity distribution with a
much weaker velocity gradient in the tail, might the tail be
dynamically isolated from the head's dark matter.

\cite{law09} include a galaxy in their OSIRIS survey, Q1700-BX710, that
appears to be a tadpole from the HST/ACS image. The redshift, $z=2.29$,
is comparable to that of our tadpoles. The diameter of the bright part,
2.2 kpc, is slightly larger than our diameters, as are the stellar
brightness, $K_S({\rm Vega})=20.23$ mag and SED mass
$M=4.4\times10^{10}\;M_\odot$. There is no regular velocity rotation in
Q1700-BX710, as the kinematics appears to be dominated by random
motions with $V_{\rm shear}/\sigma=0.2$ and $\sigma=68\pm25$ km
s$^{-1}$, calculated as the average over the dispersion for each pixel.
Law et al. state that the tail has a lower velocity dispersion than the
head and points to another galaxy 41 kpc away at the same redshift.
Unlike the tadpole galaxies in \cite{forster09} and most of those in
the present study, the tadpole in Law et al. shows good evidence for an
interaction.

\subsection{A few Odd Cases and Possible Face-on
Tadpoles} \label{sect:odd}

\begin{figure}\epsscale{1}
\plotone{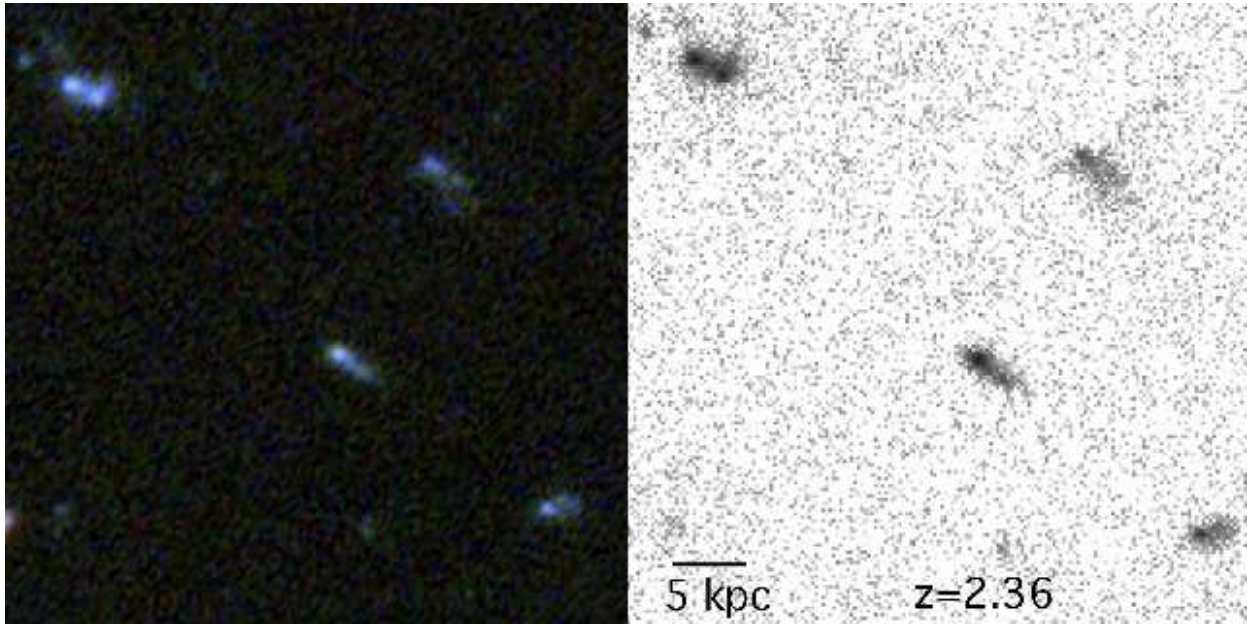} \caption{Four similar galaxies in a UDF field.
The double-clump at the top left is UDF 4699, the tadpole at the top
right is UDF 4682, the tadpole in the center is UDF 4592, and the
tadpole at the lower right is UDF 4518. The redshifts using the method
of \cite{coe06} as compiled in \cite{e07}, are 1.96, 1.72, 2.36, and
2.49. Perhaps the galaxies are closer together than these photometric
redshifts suggest. Redshifts for only two of these galaxies are in
\cite{rafel09}: 2.34 for UDF 4699 and 2.7 for UDF 4592. [Image quality
degraded for arXiv] }\label{Fig2-3tads}\end{figure}

\begin{figure}\epsscale{1}
\plotone{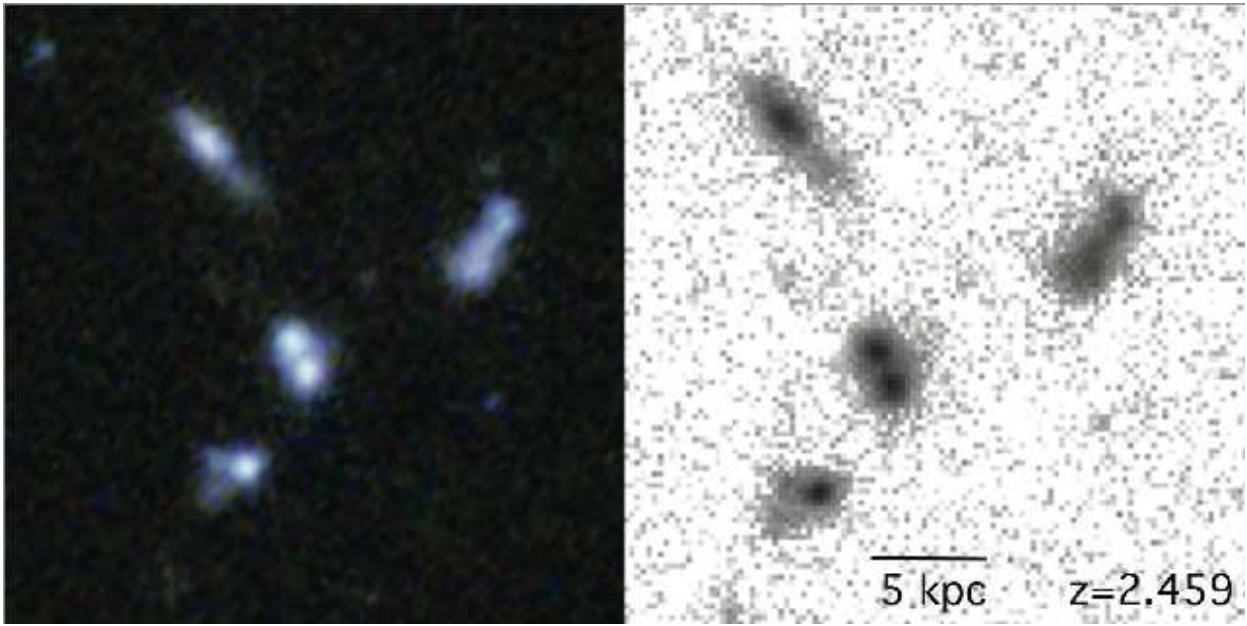} \caption{Three tadpoles and a double-clump
galaxy in the UDF field.  The galaxies are: UDF 3583 (upper left), UDF
3527 (upper right), UDF 3508 (bottom), and UDF 3513 (middle). The first
three are tadpoles and the fourth is a double-clump.  The redshifts in
\cite{e07} are 1.87, 1.73, 1.63, and 2.24, respectively. The
\cite{rafel09} catalog has two of these galaxies, UDF 3583 with
$z=2.08$ and UDF 3527 with $z=2.26$. [Image quality degraded for
arXiv]}\label{UDF1928-group}\end{figure}

Our survey of UDF tadpoles uncovered a few odd cases with suggestive
environments. Figure \ref{Fig2-3tads} shows a UDF field of view with 3
nearly-aligned tadpoles and a double-galaxy. Figure \ref{UDF1928-group}
shows another field with 4 clumpy galaxies, two of which are tadpoles.
These fields could be evidence for environment effects, such as ram
pressure stripping in dense cosmological flows.  However, recall that
tadpoles do not have a statistically significant excess or deficit of
other equally-large galaxies around them, nor do they have any
preferred orientation relative to their companions, as discussed above.
Therefore these two odd cases could be statistical flukes.

The odd cases do suggest another interpretation for tadpoles, however.
When combined with the generally irregular morphology of clumpy,
high-redshift disks, and the common occurrence of double-clump galaxies
like those in Figures \ref{Fig2-3tads} and \ref{UDF1928-group}, we get
the impression that some disks might have only one big clump, and in
the tadpole class, these single clumps are viewed at the projected
edges of the disks. This model is also consistent with the kinematical
data presented in the previous section (if it applies to our galaxies),
which suggested that the full extent of the tadpole is required for the
rotation curve to give a total dynamical mass reasonably larger than
the SED-fitted stellar mass (i.e., to account for dark matter).

\begin{figure}\epsscale{1}
\plotone{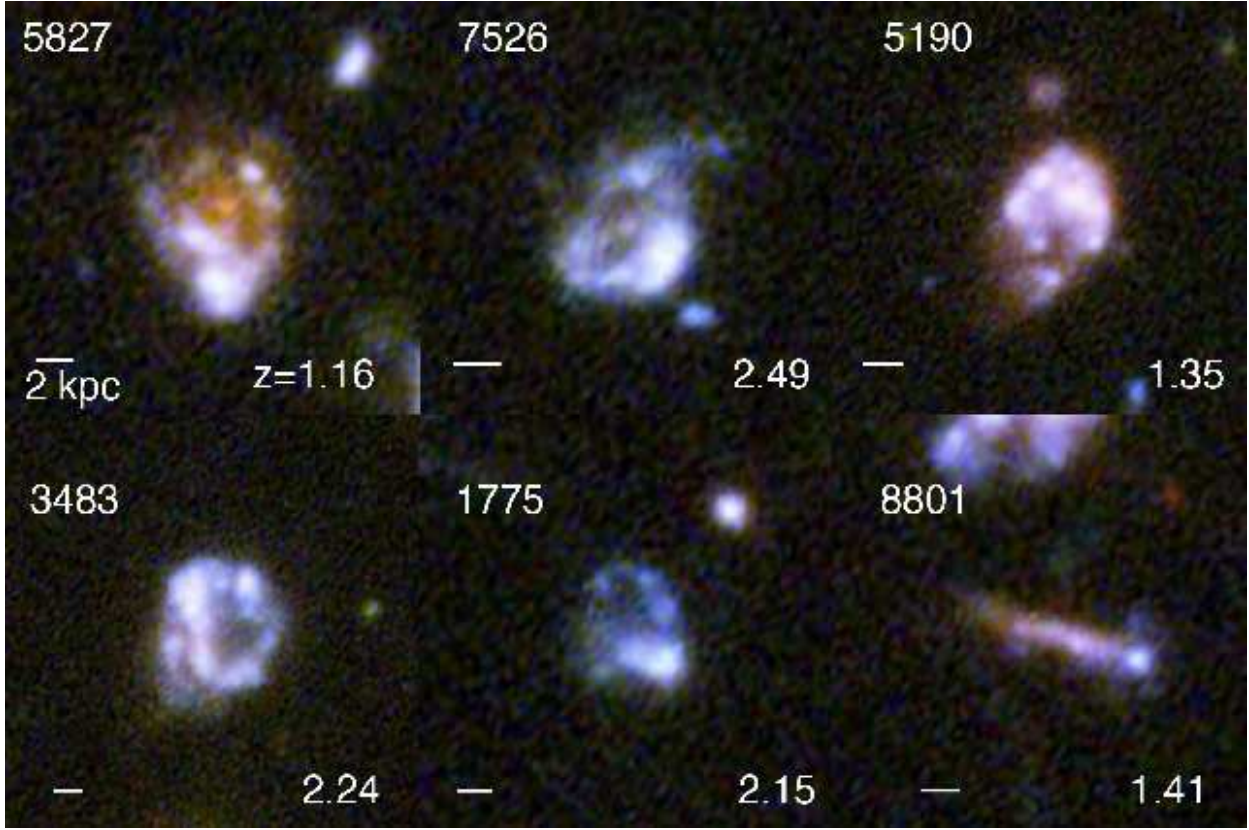} \caption{Five lop-sided galaxies in the UDF
with ring-like structure, plus one straight galaxy, probably an edge-on
disk, with a large star-forming clump at one end. The first five
galaxies might be classified as tadpoles if they were viewed in the
right orientation. The sixth might look like a tadpole if the
resolution were a little poorer. [Image quality degraded for
arXiv]}\label{6_rings}\end{figure}

Figure \ref{6_rings} shows 5 examples of clumpy disks with single,
lop-sided clumps. Some are ring-like \citep[other examples of ring-like
clump clusters are in][]{e09a}. The bottom right example is most likely
an edge-on disk with a large star forming region at one end. It is
something like a chain galaxy, but with only one clump in the chain.
The lop-sided ring clump clusters may look like tadpoles when viewed
edge-on.

A remote possibility is that the tails are the result of
supernova-driven winds consisting of driven gas and the star formation
in it, in addition to gravitationally entrained old stars.  This is
unlikely because winds leaving a galaxy (presumably the tadpole head)
should be bipolar or more isotropic than the one-sided tail. It is also
energetically unreasonable to gravitationally drag from the head the
relatively high stellar masses that we measure in the tails.

\section{Discussion}

The tadpole galaxies in the UDF have relatively low-mass heads and
tails, and they are generally young when viewed in ACS-band colors,
which are in the restframe uv.  The same is true when NICMOS colors are
added. The tail mass is comparable to or slightly larger than the head
mass, and the head ages are about the same as the tail ages, to within
the uncertainties of our model fits. We cannot tell the origin of
tadpoles from these measurements alone.

Tadpoles resemble poorly-resolved versions of tidally interacting
galaxies, such as the namesake ``Tadpole Galaxy,'' UGC 10214.
Occasionally there is a double head to support this model (Figure
\ref{3_wiggly_tads}), but this situation is rare (13\% for UDF
tadpoles).  Still, a plausible model for the origin of tadpole
structure is a galaxy merger, even though the merger remnants are not
usually visible in the tadpole heads, the tadpole tails are straighter
and more regular than most merger tails, and no excess of nearby large
galaxies is systematically seen (Sect. \ref{comp}). A $z=2.29$ tadpole
galaxy that could be a merger has a chaotic velocity field and a nearby
companion in a study by \cite{law09}(Sect.\ref{sect:dyn}).

Tadpoles also resemble comets that suggest a wind-swept origin.  They
would have to be gas-rich in this case, especially the heads, to put so
much star-forming mass in the swept-back tails. They would also have to
be moving though a dense intergalactic medium for ram pressure to have
much of an effect. Considering that the average stellar density of a
tadpole head is $\sim2\;M_\odot$ pc$^{-3}$ (Sect. \ref{sect:ma}), which
is $1.4\times10^{-22}$ gm cm$^{-3}$, the corresponding density in
hydrogen atoms, including helium, would be $\sim70$ cm$^{-3}$.  If the
gas mass in the head is comparable to the stellar mass, as in other
young galaxies \citep{tacc10}, then the gas density would be about the
same. In order to unbind a lot of this gas and move it into a tail, the
intergalactic density would have to be comparable to the head density
if the tadpole moved through this medium at about its own escape
velocity. This velocity is $\sim40$ km s$^{-1}$ from the stellar mass
and radius alone, and probably $3\times$ larger if there is $10\times$
the stellar mass in dark matter. For larger tadpole velocities, $v$,
the intergalactic density could be less by $1/v^2$ and have the same
ram pressure.

Intergalactic ram pressure is observed locally to produce a morphology
similar to what we see in tadpoles. \cite{chung09} studied the HI and
stellar distributions in Virgo Cluster galaxies. IC 3418 (their Fig. 2)
is an IBm galaxy with a uv-bright trail of stars extending 9 kpc
southeast of its main body. There is no detectable gas now but the
authors suggest that the tail formerly contained gas that was converted
into stars by the cluster pressure. Other Virgo spirals have slightly
displaced gas disks or truncated gas disks as a result of the ram
pressure from their motion through the hot intergalactic gas. NGC 4294
in Virgo has a 27 kpc long tail of HI without any stars down to 26 mag
arcsec$^2$. Its neighbor, NGC 4299, has a similar starless HI tail
parallel to the one in NGC 4294. These tails are probably from ram
pressure, but the two galaxies also show signs of an interaction.

For the tadpoles in the UDF, the tails are much more massive in young
stars than they are for the tadpole in Virgo. The intergalactic density
should be larger for the UDF tadpoles if there are active cosmological
gas inflows to nearby galaxies. The UDF tadpoles could be going through
these dense flows. The velocity of the UDF tadpoles through the
intergalactic medium should be smaller than the infall velocities in
Virgo because the Virgo potential well is deep now; massive galaxy
clusters were not well developed at $z\sim2$. Whether the ram pressure
was greater for the UDF tadpoles than the Virgo tadpole cannot be
determined yet, but even at the same ram pressure, the high
intergalactic density in the UDF should have a strong gas-stripping
effect on the tadpoles. Numerical simulations of young galaxies moving
through cosmological flows could test this. If a low mass galaxy with a
high gas fraction moves through a dense, cold-flow region, then we
predict a large fraction of the gas will get drawn into a tail. Star
formation should occur in this tail, and because the gas fraction is
high, a significant fraction of old stars in the original galaxy could
get drawn into the tail as well.

Tadpoles could also be edge-on disks with a single, large, star-forming
clump. If the clump is near the center, we might say the galaxy is
normal, edge-on, and has a bulge, or we might call it a chain galaxy if
the clump is very blue. When the clump is near the edge of the disk,
however, it should look like a tadpole when viewed edge-on from the
right orientation. Some examples of lop-sided, single-clump galaxies
were shown in Section \ref{sect:odd}. This third model for tadpoles
suggests they are single disks with a common halo of dark matter. This
is the most logical explanation for the three tadpoles studied by
\cite{forster09} that have position-velocity distributions with
near-constant velocity gradients.

\section{Conclusions}

Tadpole galaxies as a morphological class could be a mixture of several
types: (1) mergers, which should show double-heads if resolved
properly, chaotic internal motions, and tails with peculiar velocities
relative to the head rotation curves, (2) ram-pressure stripped heads
and their debris tails that are interacting with a dense intergalactic
gas or cosmological gas flow, (3) ram-pressure induced star formation
at the edge of a low surface brightness galaxy disk, viewed edge-on,
and (4) naturally lop-sided clumpy disks that are viewed edge-on.

Tadpole heads in the UDF have photometric masses from the rest-frame uv
and visible passbands that are in the range of $10^6-10^9\;M_\odot$,
increasing with redshift because of selection effects on size, surface
brightness, and rest wavelength. The masses are comparable to the
masses of bulge-like objects in other UDF clumpy galaxies, and also
comparable to the largest clumps in clump-clusters and chains. Head
ages span a wide range around and below 1 Gyr, but are poorly
constrained by the SEDs considering the uncertainty with internal
extinction.

Acknowledgmenta: We are grateful to the referee for helpful comments.
DME thanks Vassar College for support from the astronomy publication
fund.

{}

\clearpage
%fig1

\clearpage
%fig2

\clearpage
%fig3

\clearpage
%fig4

\clearpage
%fig5

\clearpage
%fig6

\clearpage
%fig7

\clearpage
%fig8

\clearpage
%fig9

\clearpage
%fig10

\clearpage
%fig11

\clearpage
%fig12

\clearpage
%fig13

\clearpage
%fig14

\clearpage
%fig15

\clearpage
%fig16

\end{document}